%% file: main.tex
\documentclass[10pt,journal,twoside]{IEEEtran}

\usepackage{amsmath,amssymb,amsfonts}
\usepackage{mathtools}
\usepackage{bm}
\usepackage{cite}
\usepackage{graphicx}
\usepackage{booktabs}
\usepackage{algorithmic}

\newtheorem{theorem}{Theorem}
\newtheorem{proposition}{Proposition}
\newtheorem{corollary}{Corollary}
\newtheorem{lemma}{Lemma}
\newtheorem{assumption}{Assumption}
\newtheorem{definition}{Definition}

\newcommand{\R}{\mathbb{R}}
\newcommand{\Hcal}{\mathcal{H}}
\newcommand{\Ycal}{\mathcal{Y}}

\newcommand{\dist}{\operatorname{dist}}
\newcommand{\rank}{\operatorname{rank}}
\newcommand{\im}{\operatorname{Im}}
\newcommand{\Span}{\operatorname{span}}

\title{Singular Sampling in Nonlinear Observability: Higher-Order Minimax Laws and Timing Crossovers}

\author{Jerzy~Baranowski,~\IEEEmembership{Senior Member,~IEEE}%
\thanks{Jerzy Baranowski is with the Department of Automatic Control and Robotics, AGH University of Krakow, 30--059 Krakow, Poland (e-mail: jb@agh.edu.pl).}}

\begin{document}

\maketitle

\begin{abstract}
We study deterministic bounded-noise state reconstruction for nonlinear systems when the measurement schedule drives the sampled-output map through a rank-deficient configuration. Local reconstruction is governed by the leading pairwise separation order and its corresponding higher-order modulus, which together determine the leading minimax noise-to-state law. For corank-one cubic singular schedules, including overdetermined sampled maps, we derive the exact physical $\ell_\infty$ cubic modulus through quotient geometry in $\mathbb{R}^q/\operatorname{Im}DF$ and show how the required finite jet is computed from trajectory sensitivities. We then prove a uniform signed timing crossover: a transversal timing perturbation has transition scale $\varepsilon=O(\delta^{2/3})$, linking the regular, cubic-singular, and fold regimes. A damped pendulum provides closed-form crossover coefficients and exact nonlinear minimax validation, while a four-state downward cart--pole verifies the rectangular theory. The results show that measurement timing can change not only reconstruction conditioning but the exponent governing bounded-noise state uncertainty.
\end{abstract}


\input{sections/01_introduction}
\input{sections/02_problem}
\input{sections/03_cubic}
\input{sections/04_crossover}
\input{sections/05_benchmarks}
\input{sections/06_conclusion}

\section*{Acknowledgment}
The author used OpenAI ChatGPT during preparation of this article for editorial and organizational assistance, language refinement, and assistance in developing and checking mathematical and computational drafts. This assistance affected the presentation throughout the manuscript. All mathematical derivations, numerical results, citations, scientific claims, and final text were reviewed and validated by the author, who assumes full responsibility for the content.

\appendices
\input{sections/appendix_proofs}

\bibliographystyle{IEEEtran}
\bibliography{references}

\begin{IEEEbiography}[{\includegraphics[width=1in,height=1.25in,clip,keepaspectratio]{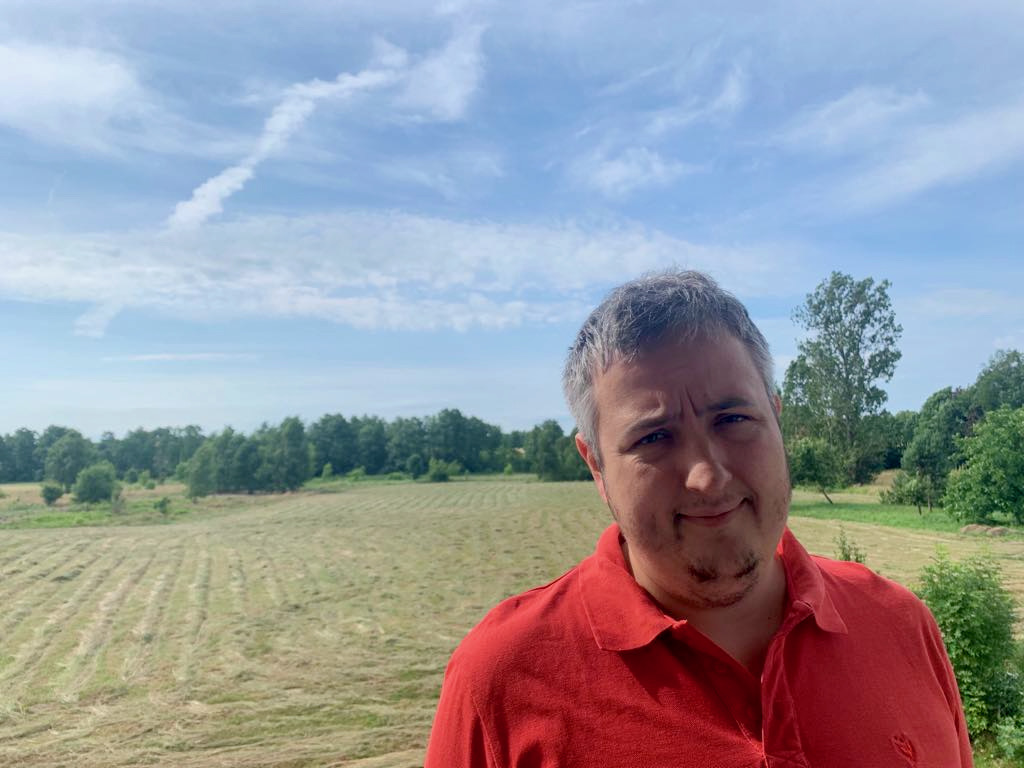}}]{Jerzy Baranowski}
(M'12--SM'21) received the M.Sc. and Ph.D. degrees in automatic control from AGH University of Science and Technology, Krakow, Poland, in 2006 and 2010, respectively, and the habilitation degree in automatic control and robotics in 2017. He is currently a Full Professor with the Department of Automatic Control and Robotics, AGH University of Krakow, where he heads the Laboratory of Computer Science in Control and Management. His research interests include control theory, process diagnostics, statistical modeling, Bayesian data analysis, numerical methods, and machine learning. He is a member of the IEEE Control Systems Society.
\end{IEEEbiography}

\end{document}

%% file: sections/01_introduction.tex
\section{Introduction}
\label{sec:introduction}

Observability of nonlinear systems is classically formulated through injectivity or immersion properties of an observation map, with differential rank conditions providing local tests under regularity assumptions \cite{Aeyels1981,Ammar2020}. Quantitative observability measures beyond a binary rank test have also been developed for nonlinear systems, including empirical- and variational-Gramian approaches and Koopman-based characterizations \cite{Zeng2018,Mesbahi2021,KazmaTaha2024}. Sampling adds a second geometric layer: even when the continuous-time system is observable, particular measurement times can make the finite sampled-output map rank deficient. Preservation and genericity of observability under sampling are established topics \cite{Ammar2004,Ammar2016,Ammar2018,Ammar2020}, while nonlinear observer theory and sampled-output observer and estimator design form a substantial parallel literature \cite{BernardAndrieuAstolfi2022,Dorleans2011,Mazenc2022,Habibi2023,Menard2023}. What a rank-deficient sampling schedule implies for the local information limit of \emph{bounded-noise state reconstruction}, however, is more delicate. A singular Jacobian can coexist with local separation by the exact nonlinear sampled-output map, so a first-order conditioning measure can degenerate while higher-order information remains available. This setting is distinct from genuine indistinguishability, where different state trajectories generate the same output and state reconstruction is intrinsically set-valued \cite{BernardMaghenem2024}.

Fractional inverse-stability estimates are well known in variational analysis through H\"older metric subregularity and related error-bound theory \cite{LiMordukhovich2012,Cibulka2018}. Singularity theory likewise provides finite-jet descriptions of degenerate maps, including quantitative variants motivated by finite-accuracy computations \cite{GolubitskyGuillemin1973,Yomdin2005}. In noisy inverse problems, singularity-based error amplification has been studied explicitly, for example for Prony-type reconstruction near colliding nodes \cite{Goldman2018}. Bounded-error nonlinear state estimation and set-membership methods form another mature literature, with recent developments for constrained nonlinear systems under unknown-but-bounded uncertainty \cite{Scholte2003,Alamo2008,dePaula2024}. Finally, selection of measurement times is a classical design variable, from early sampling-instant optimization to nonlinear experimental design and estimation/control schemes with optimized measurement times \cite{Troch1973,BanksRehm2014,Jost2017}. The objective here is therefore not to introduce any of these surrounding concepts in isolation.

This paper connects them at the level of a nonlinear dynamical sampled-output map and asks a timing-specific minimax question. For a fixed measurement schedule $\tau$, we consider arbitrary estimators under bounded measurement error and characterize the local worst-case reconstruction law through the pairwise separation geometry of the exact sampled-output map. At regular schedules this reduces to the first-order physical minimum gain. At corank-one singular schedules, the leading separating information can instead be cubic, producing a $\delta^{1/3}$ reconstruction law with a computable coefficient. Varying only the measurement times can therefore change not merely a conditioning constant but the exponent of the noise-to-state relation.

The terminology used below reflects this local geometry. A schedule is \emph{singular} at a nominal state when the Jacobian of its sampled-output map loses column rank; its \emph{corank} is the dimension of the resulting right nullspace. A corank-one singular schedule is called \emph{cubic no-fold} when the linearly invisible state direction is still separated by the exact nonlinear map and the first nonvanishing reduced pairwise term is of odd order three: the scalar normal picture is $s\mapsto \gamma s^3$, rather than the fold geometry $s\mapsto s^2$. Thus, \emph{cubic} refers to the leading separating order, whereas \emph{no-fold} emphasizes that local state separation is retained despite first-order rank loss. By a \emph{timing crossover} we mean the joint asymptotic transition obtained when the schedule is detuned through such a singular timing while the measurement-noise level tends to zero.

The contributions are as follows.
\begin{itemize}
    \item We formulate a schedule-dependent \emph{robust reconstruction order} $k(\tau)$ and higher-order observability modulus $\mu^{(k)}(\tau)$ from deterministic pairwise separation geometry. When the corresponding finite-type law holds, these quantities determine the local minimax asymptotic $R^\star(\delta)\sim 2^{1/k-1}[\mu^{(k)}]^{-1/k}\delta^{1/k}$.
    \item For corank-one cubic schedules, including overdetermined maps $F:\R^n\to\R^q$ with $q\ge n$, we derive an intrinsic quotient-space formulation in $\R^q/\im DF$. The resulting cubic modulus is computed from a finite-dimensional homogeneous problem in the physical state and output norms, with a directly checkable sufficient condition based on second and third derivatives of the sampled-output map.
    \item We express the required derivatives through first-, second-, and third-order trajectory sensitivity equations. The singular-schedule test is therefore a system-level computation and does not require construction of a nonlinear inverse observation map.
    \item We prove a uniform signed crossover theorem when a timing parameter passes transversally through a cubic singular schedule. The transition layer has scale $\varepsilon=O(\delta^{2/3})$. In the scalar strict-cubic case, the complete scaled profile is the largest positive root of a cubic equation and links a regular side, the cubic singular point, and a fold side with a distinct ambiguity mechanism.
\end{itemize}

Two nonlinear mechanical examples illustrate complementary aspects of the theory. A damped pendulum admits closed-form singular timing, cubic modulus, and crossover coefficients, allowing direct comparison with exact nonlinear minimax computations. A four-state downward cart--pole produces a genuinely rectangular $4\to6$ sampled-output map and demonstrates the trajectory-sensitivity and quotient-space calculations beyond the two-state setting.

The paper is organized as follows. Section~\ref{sec:problem} introduces the bounded-noise reconstruction problem and finite-type quantities. Section~\ref{sec:cubic} develops the rectangular cubic corank-one theory and its sensitivity realization. Section~\ref{sec:crossover} derives the uniform timing crossover. Section~\ref{sec:benchmarks} presents the nonlinear benchmarks, and Section~\ref{sec:conclusion} summarizes the implications for measurement-time design.

%% file: sections/02_problem.tex
\section{Bounded-Noise Sampled Reconstruction}
\label{sec:problem}

Consider
\begin{equation}
    \dot x(t)=f(t,x(t),u(t)),
    \qquad
    y(t)=h(t,x(t))+v(t),
\end{equation}
where the input $u$ is known and the measurement disturbance is bounded. For a schedule
\begin{equation}
    \tau=(t_1,\ldots,t_m),
\end{equation}
define the sampled-output map
\begin{equation}
    \Hcal_\tau(x_0)
    =
    \begin{bmatrix}
        h(t_1,x(t_1;x_0))\\
        \vdots\\
        h(t_m,x(t_m;x_0))
    \end{bmatrix}
    \in\R^q.
\end{equation}
At a nominal state $\bar x$, write
\begin{equation}
    J_\tau(\bar x)=D\Hcal_\tau(\bar x).
\end{equation}
We call the schedule \emph{regular} at $\bar x$ if $J_\tau(\bar x)$ has full column rank, and \emph{singular} otherwise. Its corank is
\begin{equation}
    \operatorname{corank}J_\tau(\bar x)
    =n-\rank J_\tau(\bar x)
    =\dim\ker J_\tau(\bar x).
\end{equation}
Thus, \emph{singular sampling} refers here to a first-order degeneracy of the finite sampled-output map for a particular schedule and nominal state. It does not by itself imply loss of nonlinear observability of the continuous-time system, nor noninjectivity of the exact sampled-output map.

The measured vector is
\begin{equation}
    z=\Hcal_\tau(x_0)+e,
    \qquad
    \|e\|_\infty\le\delta.
\end{equation}
Weighted physical state and output norms can be reduced to this form by diagonal scaling; throughout the theoretical development the scaled coordinates are understood as fixed.

Let $X\subset\R^n$ be a compact local state set. For an arbitrary estimator $\widehat x:\R^q\to\R^n$, define
\begin{equation}
    \mathcal R_X(\widehat x,F,\delta)
    =
    \sup_{\substack{x\in X\\\|e\|_\infty\le\delta}}
    \|\widehat x(F(x)+e)-x\|_\infty,
\end{equation}
and
\begin{equation}
    R_X^\star(F,\delta)
    =\inf_{\widehat x}\mathcal R_X(\widehat x,F,\delta).
\end{equation}

\begin{proposition}[Pairwise minimax geometry]
\label{prop:pairwise}
For continuous $F:X\to\R^q$, define
\begin{equation}
\begin{aligned}
    d_F(\delta)
    =\sup\{\, &\|x-x'\|_\infty:
    x,x'\in X,\\
    &\|F(x)-F(x')\|_\infty\le2\delta\,\}.
\end{aligned}
\end{equation}
Then
\begin{equation}
    \boxed{
    R_X^\star(F,\delta)=\frac12d_F(\delta).}
\end{equation}
\end{proposition}

Proposition~\ref{prop:pairwise} is an optimal-recovery statement specialized to the $\ell_\infty$ geometry used here; worst-case recovery from inaccurate information has a classical theory \cite{MicchelliRivlin1977}. Its role here is foundational rather than a novelty claim: it reduces arbitrary-estimator minimax analysis to the separation properties of the sampled observation map.

For a nominal state $\bar x$ and an integer $k\ge1$, define the pointed $k$th-order separation quotient by
\begin{equation}
\label{eq:pointed-modulus}
\mu^{(k)}(F,\bar x)
:=
\lim_{\rho\downarrow0}
\inf_{\substack{x,x'\in B_\infty(\bar x,\rho)\\x\ne x'}}
\frac{\|F(x)-F(x')\|_\infty}
{\|x-x'\|_\infty^k},
\end{equation}
where the limit exists as an extended real number because the infima are monotone under neighborhood restriction.

\begin{definition}[Robust reconstruction order]
\label{def:order}
The map $F$ has \emph{robust reconstruction order} $k$ at $\bar x$ if
\begin{equation}
    0<\mu^{(k)}(F,\bar x)<\infty.
\end{equation}
The corresponding $\mu^{(k)}(F,\bar x)$ is the \emph{higher-order observability modulus}.
\end{definition}

For a fixed local neighborhood $X_\rho=B_\infty(\bar x,\rho)$, let
\begin{equation}
    \omega_{F,\rho}(r)
    =
    \inf_{\substack{x,x'\in X_\rho\\
    \|x-x'\|_\infty\ge r}}
    \|F(x)-F(x')\|_\infty.
\end{equation}
Whenever, on every sufficiently small fixed neighborhood, the sharper pairwise law
\begin{equation}
    \omega_{F,\rho}(r)
    =\mu^{(k)}(F,\bar x)r^k+o(r^k),
    \qquad r\downarrow0,
\end{equation}
holds with the same leading coefficient, Proposition~\ref{prop:pairwise} yields the local minimax asymptotic
\begin{equation}
\label{eq:minimax-finite-type}
    \boxed{
    R_{\rm loc}^\star(F,\delta)
    =
    2^{1/k-1}
    [\mu^{(k)}(F,\bar x)]^{-1/k}
    \delta^{1/k}
    +o(\delta^{1/k}).}
\end{equation}

For a regular map, $J=DF(\bar x)$ has full column rank and
\begin{equation}
    \mu^{(1)}(F,\bar x)
    =
    \mu_\infty(J)
    :=
    \min_{\|d\|_\infty=1}\|Jd\|_\infty.
\end{equation}
Thus the usual first-order minimum gain is the order-one member of the same hierarchy. The issue studied below is what replaces it when $J$ loses one rank direction but the exact nonlinear map remains locally separating.

%% file: sections/03_cubic.tex
\section{Cubic Corank-One Sampling Schedules}
\label{sec:cubic}

Let $F:\R^n\to\R^q$, $q\ge n$, be $C^4$ near the origin, with $F(0)=0$, and set
\begin{equation}
    J=DF(0),
    \qquad
    \rank J=n-1.
\end{equation}
Let $v$ span $\ker J$. The overdetermined case differs from the square case because $\ker J^T$ is no longer one-dimensional. The intrinsic nonlinear information invisible to first order therefore lives in the quotient
\begin{equation}
    \Ycal_J
    :=\R^q/\im J.
\end{equation}
Denote by $\pi_J$ the quotient projection. The physical output $\ell_\infty$ norm induces
\begin{equation}
\label{eq:quotient-norm}
    \|\pi_J(y)\|_{J,\infty}
    :=
    \dist_\infty(y,\im J)
    =
    \min_u\|y-Ju\|_\infty.
\end{equation}
By duality,
\begin{equation}
\label{eq:quotient-dual}
    \|\pi_J(y)\|_{J,\infty}
    =
    \max_{\substack{\lambda\in\ker J^T\\
    \|\lambda\|_1\le1}}
    |\lambda^Ty|.
\end{equation}

Choose a row map $P\in\R^{(n-1)\times q}$ and a covector $\zeta$ satisfying $\zeta^Tv=1$ such that
\begin{equation}
    T=
    \begin{bmatrix}
        PJ\\
        \zeta^T
    \end{bmatrix}
\end{equation}
is nonsingular. The local coordinate map
\begin{equation}
    \Theta(x)
    =
    \begin{bmatrix}
        PF(x)\\
        \zeta^Tx
    \end{bmatrix}
    =
    \begin{bmatrix}w\\z\end{bmatrix}
\end{equation}
is a local diffeomorphism. Let $x=X(w,z)$ denote its inverse and define the quotient-valued reduced output
\begin{equation}
    Q(w,z)
    =\pi_J(F(X(w,z))).
\end{equation}
The regular coordinates satisfy the exact identity $PF(X(w,z))=w$.

The first possible obstruction to a uniform cubic mechanism is a mixed second-order term.

\begin{lemma}[Mixed second-order obstruction]
\label{lem:mixed-obstruction}
A necessary condition for a neighborhood cubic no-fold geometry is
\begin{equation}
\label{eq:mixed-obstruction}
    \boxed{
    D^2F(0)[v,p]\in\im J
    \qquad\forall p\in\R^n.}
\end{equation}
Equivalently,
\begin{equation}
    \lambda^T D^2F(0)[v,p]=0
\end{equation}
for every $p$ and every $\lambda\in\ker J^T$.
\end{lemma}

Under \eqref{eq:mixed-obstruction}, the reduced derivative has no linear term. Write
\begin{equation}
\label{eq:Qz-cubic}
    Q_z(\xi)
    =R_2(\xi)+O(\|\xi\|^3),
    \qquad
    \xi=(w,z),
\end{equation}
where $R_2$ is a homogeneous quotient-valued quadratic map. Let $\mathcal M_J$ be its symmetric bilinear polarization, so that
\begin{equation}
    R_2(\xi)=\frac12\mathcal M_J(\xi,\xi).
\end{equation}
Define the cubic finite-difference jet
\begin{equation}
\label{eq:Gamma3}
\begin{aligned}
    \Gamma_3(c)
    &:=\int_{-1/2}^{1/2}R_2(c+se_n)\,ds\\
    &=\frac12\mathcal M_J(c,c)
    +\frac1{24}\mathcal M_J(e_n,e_n).
\end{aligned}
\end{equation}

Some systems possess exact common-shift symmetries that cannot affect any pairwise reconstruction problem.

\begin{definition}[Exact common-shift symmetry]
\label{def:common-shift}
A linear subspace $\mathcal S\subset\R^n$ is an exact common-shift symmetry if, for all sufficiently small $x$ and $s\in\mathcal S$,
\begin{equation}
\label{eq:common-shift}
    F(x+s)=F(x)+L(s),
\end{equation}
where $L:\mathcal S\to\R^q$ is independent of $x$. Then
\begin{equation}
    F(x+s)-F(x'+s)=F(x)-F(x')
\end{equation}
for every pair $x,x'$.
\end{definition}

A common shift also leaves $x-x'$ unchanged. Hence midpoint coordinates along $\mathcal S$ are gauge variables for the pairwise separation problem. In reduced coordinates define the linearized symmetry subspace
\begin{equation}
    \mathcal S_\xi
    :=D\Theta(0)\mathcal S
    =DX(0)^{-1}\mathcal S,
\end{equation}
and choose any complement
\begin{equation}
    \R^n=\mathcal S_\xi\oplus\mathcal N_\xi.
\end{equation}
If no nontrivial exact symmetry is present, take $\mathcal S_\xi=\{0\}$ and $\mathcal N_\xi=\R^n$. Define
\begin{equation}
\label{eq:chi3rect}
    \chi_3^{\rm rect}
    :=
    \inf_{c\in\mathcal N_\xi}
    \|\Gamma_3(c)\|_{J,\infty}.
\end{equation}
The value is independent of the chosen transversal complement because all such complements represent the same pairwise-difference classes modulo the exact symmetry.

\begin{assumption}[Cubic no-fold coercivity]
\label{ass:cubic-coercivity}
The effective cubic jet satisfies
\begin{equation}
    \chi_3^{\rm rect}>0
\end{equation}
and is coercive on the effective midpoint space:
\begin{equation}
    \|\Gamma_3(c)\|_{J,\infty}\to\infty
    \quad\text{as}\quad
    \|c\|\to\infty,
    \qquad c\in\mathcal N_\xi.
\end{equation}
\end{assumption}

\begin{theorem}[Exact rectangular cubic modulus]
\label{thm:rectangular-cubic}
Suppose that $F$ is $C^4$, $\rank DF(0)=n-1$, the mixed obstruction \eqref{eq:mixed-obstruction} vanishes, and Assumption~\ref{ass:cubic-coercivity} holds after removal of any exact common-shift symmetry in Definition~\ref{def:common-shift}. Then $F$ has robust reconstruction order three and
\begin{equation}
\label{eq:mu3-rectangular}
    \boxed{
    \mu^{(3)}(F,0)
    =
    \frac{\chi_3^{\rm rect}}
    {\|v\|_\infty^3}.}
\end{equation}
Consequently,
\begin{equation}
\label{eq:risk-cubic-rectangular}
    \boxed{
    R_{\rm loc}^\star(F,\delta)
    =
    2^{-2/3}
    \left(
    \frac{\|v\|_\infty^3}
    {\chi_3^{\rm rect}}
    \right)^{1/3}
    \delta^{1/3}
    +o(\delta^{1/3}).}
\end{equation}
\end{theorem}

\begin{figure*}[t]
    \centering
    \includegraphics{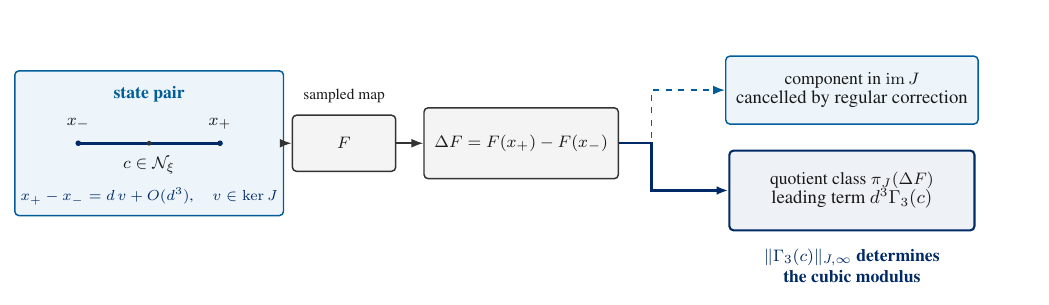}
    \caption{Cubic corank-one geometry. The state pair is separated predominantly along $v\in\ker J$. Third-order corrections in regular state directions cancel the output component in $\im J$, whereas the quotient class $\pi_J(\Delta F)$ remains and is governed by the effective cubic jet $\Gamma_3(c)$.}
    \label{fig:cubic-geometry}
\end{figure*}

The proof is given in Appendix~\ref{app:rectangular-proof}. The key point is that $O(r^3)$ corrections in regular state directions can cancel exactly the component of the cubic output increment lying in $\im J$, while they cannot alter its quotient class. Hence the leading physical output norm is precisely the quotient distance in \eqref{eq:quotient-norm}, not a conservative bound. Exact common-shift symmetries are removed before the scaled-midpoint coercivity argument because they leave pairwise state and output differences invariant identically. Figure~\ref{fig:cubic-geometry} summarizes this geometry.

For computation, it is useful to contract the quotient-valued jet by left-null covectors. For $\lambda\in\ker J^T$, let $M_\lambda$ be the scalar symmetric matrix obtained by applying $\lambda^T$ to a physical representative of $\mathcal M_J$.

\begin{corollary}[Scalar cubic witness]
\label{cor:scalar-witness}
Suppose there exist $\lambda\in\ker J^T\setminus\{0\}$ and $\sigma\in\{-1,1\}$ such that
\begin{equation}
    c^T(\sigma M_\lambda)c>0
    \qquad
    \forall c\in\mathcal N_\xi\setminus\{0\},
\end{equation}
and
\begin{equation}
    e_n^T(\sigma M_\lambda)e_n>0.
\end{equation}
Then
\begin{equation}
    \sigma\lambda^T\Gamma_3(c)
    =
    \frac12c^T(\sigma M_\lambda)c
    +\frac1{24}e_n^T(\sigma M_\lambda)e_n
\end{equation}
is positive and coercive on $\mathcal N_\xi$. By \eqref{eq:quotient-dual}, Assumption~\ref{ass:cubic-coercivity} follows and Theorem~\ref{thm:rectangular-cubic} applies. In the absence of exact common-shift symmetries, the sufficient condition reduces to
\begin{equation}
\label{eq:Mlambda-pd}
    \boxed{\sigma M_\lambda\succ0.}
\end{equation}
\end{corollary}

When $q=n$, $\ker J^T=\Span\{\lambda\}$ and the quotient is one-dimensional. In the strict cubic case, \eqref{eq:mu3-rectangular} reduces to
\begin{equation}
\label{eq:mu3-square}
    \boxed{
    \mu^{(3)}
    =
    \frac{|\kappa_3|}
    {24\|\lambda\|_1\|v\|_\infty^3}.}
\end{equation}

\subsection{Direct computation from trajectory sensitivities}

For $F=\Hcal_\tau$, the tensors required above are obtained directly from variational equations. Along a nominal trajectory $\bar x(t)$, let
\begin{equation}
    A(t)=D_xf(t,\bar x(t),u(t)),
\end{equation}
and let $S_1[a]$ denote the first state sensitivity in direction $a$. Then
\begin{equation}
    \dot S_1[a]=A(t)S_1[a],
\end{equation}
with initial condition
\begin{equation}
    S_1[a](t_0)=a.
\end{equation}
The second and third sensitivities satisfy
\begin{align}
    \dot S_2[a,b]
    &=A S_2[a,b]
    +D_x^2f[S_1[a],S_1[b]],\\
    \dot S_3[a,b,c]
    &=A S_3[a,b,c]
    +D_x^3f[S_1[a],S_1[b],S_1[c]]\nonumber\\
    &\quad
    +D_x^2f[S_2[a,b],S_1[c]]
    +D_x^2f[S_2[a,c],S_1[b]]\nonumber\\
    &\quad
    +D_x^2f[S_2[b,c],S_1[a]],
\end{align}
with
\begin{equation}
    S_2[a,b](t_0)=0,
    \qquad
    S_3[a,b,c](t_0)=0.
\end{equation}
Combining these state sensitivities with derivatives of $h$ at the sample times gives $D\Hcal_\tau$, $D^2\Hcal_\tau$, and $D^3\Hcal_\tau$.

Thus a fixed schedule is analyzed by: (i) trajectory and sensitivity integration; (ii) nullspace computation; (iii) the mixed-obstruction test; (iv) construction of $\mathcal M_J$ or a scalar witness $M_\lambda$; and (v) the finite-dimensional optimization \eqref{eq:chi3rect}. No nonlinear inverse observation map is constructed numerically.

%% file: sections/04_crossover.tex
\section{Timing Through a Cubic Singular Schedule}
\label{sec:crossover}

The preceding section characterizes a fixed singular schedule. We now let the measurement times vary through such a schedule and study the joint limit in which timing detuning and measurement noise vanish together.

Let $F_\varepsilon$ be a $C^4$ one-parameter family of sampled-output maps with a cubic corank-one singularity at $\varepsilon=0$. The parameter $\varepsilon$ may be one component or one smooth direction in the measurement schedule. Use the fixed quotient
\begin{equation}
    \Ycal_0
    =
    \R^q/\im J_0,
    \qquad
    J_0=D_xF_0(0),
\end{equation}
and the parameter-dependent regular coordinates from Section~\ref{sec:cubic}. Under a transversal unfolding,
\begin{equation}
\label{eq:crossover-local-expansion}
    Q_{\varepsilon,z}(\xi)
    =
    \varepsilon A
    +R_2(\xi)
    +E(\varepsilon,\xi),
\end{equation}
where $A\in\Ycal_0$ is nonzero and
\begin{equation}
    \|E(\varepsilon,\xi)\|_{0,\infty}
    \le
    C\left(
    \varepsilon^2
    +|\varepsilon|\|\xi\|
    +\|\xi\|^3
    \right).
\end{equation}
If an exact common-shift symmetry is removed from the midpoint problem, we assume that the same physical symmetry is exact for all sufficiently small $\varepsilon$; the variable $c$ below then ranges over the corresponding effective midpoint space $\mathcal N_\xi$. In the absence of such a symmetry, set $\mathcal N_\xi=\R^n$.

Balancing the linear unfolding term $\varepsilon d$ with the cubic singular term $d^3$ gives
\begin{equation}
    |\varepsilon|=O(d^2).
\end{equation}
Since an indistinguishable cubic pair has $d=O(\delta^{1/3})$, the timing/noise transition layer is
\begin{equation}
\label{eq:transition-scaling}
    \boxed{
    \varepsilon=\eta\delta^{2/3}.}
\end{equation}

For $R\ge0$, define
\begin{equation}
\label{eq:Geta}
    G_\eta(R)
    =
    \inf_{c\in\mathcal N_\xi}
    \left\|
    \eta R A
    +R^3\Gamma_3(c)
    \right\|_{0,\infty},
\end{equation}
and let
\begin{equation}
\label{eq:Rstar-eta}
    R_\star(\eta)
    =
    \sup\{R\ge0:G_\eta(R)\le2\}.
\end{equation}
Set
\begin{equation}
\label{eq:Deta}
    D(\eta)
    =
    \|v\|_\infty R_\star(\eta).
\end{equation}
The use of the largest feasible $R$ is essential: on the fold side the output-separation curve need not be monotone and can cross the noise threshold more than once.

The crossover is a local statement. Accordingly, let
\begin{equation}
    X_\delta=B_\infty(0,\rho(\delta)),
\end{equation}
where
\begin{equation}
    \rho(\delta)\downarrow0,
    \qquad
    \delta^{1/3}=o(\rho(\delta)).
\end{equation}
This localization retains all $O(\delta^{1/3})$ transition-layer pairs while excluding unrelated global ambiguities.

\begin{theorem}[Uniform signed timing crossover]
\label{thm:crossover}
Assume the cubic hypotheses of Theorem~\ref{thm:rectangular-cubic} and the uniform transversal expansion \eqref{eq:crossover-local-expansion}. Let $K\subset\R$ be compact and suppose, for every $\eta\in K$, that
\begin{equation}
\label{eq:boundary-approach}
    \sup\{R:G_\eta(R)<2\}
    =
    \sup\{R:G_\eta(R)\le2\}
    =R_\star(\eta),
\end{equation}
and that $R_\star$ is continuous on $K$. Then
\begin{equation}
\label{eq:crossover-theorem}
    \boxed{
    R_{X_\delta}^\star
    \left(
    F_{\eta\delta^{2/3}},\delta
    \right)
    =
    \frac12D(\eta)\delta^{1/3}
    +o_K(\delta^{1/3})}
\end{equation}
uniformly for $\eta\in K$.
\end{theorem}

The proof is given in Appendix~\ref{app:crossover-proof}. It has two compactness steps. First, every indistinguishable pair in the transition layer has state separation $O(\delta^{1/3})$. Second, after removing any family-wide exact common-shift symmetry, its scaled midpoint remains bounded. The normalized quotient output then converges to \eqref{eq:Geta}, and the minimax diameter follows from Proposition~\ref{prop:pairwise}.

A particularly transparent expression results when the quotient is one-dimensional.

\begin{corollary}[Scalar strict-cubic profile]
\label{cor:scalar-crossover}
Suppose that the quotient coordinate is normalized by its physical norm and, on the effective midpoint space,
\begin{equation}
    \Gamma_3(c)
    =
    \gamma_0+\frac12c^TMc,
    \qquad
    \gamma_0>0,
    \qquad
    M\succ0.
\end{equation}
Let $a\ne0$ be the scalar transversal coefficient corresponding to $A$. Then $D(\eta)=\|v\|_\infty R_\eta$, where $R_\eta$ is the largest positive root of
\begin{equation}
\label{eq:cubic-profile-root}
    \boxed{
    \gamma_0R^3+a\eta R=2.}
\end{equation}
The boundary condition \eqref{eq:boundary-approach} and continuity required in Theorem~\ref{thm:crossover} then hold automatically on compact $\eta$-sets. Consequently,
\begin{equation}
    R_{X_\delta}^\star
    =
    \frac{\|v\|_\infty}{2}
    R_\eta\delta^{1/3}
    +o_K(\delta^{1/3}).
\end{equation}
\end{corollary}

Equation~\eqref{eq:cubic-profile-root} exposes a signed asymmetry that is hidden by a simple matching argument based only on $|\varepsilon|$. At the level of the crossover profile, if $a\eta\to+\infty$, then
\begin{equation}
    R_\eta\sim\frac{2}{a\eta},
\end{equation}
which matches the ordinary regular law
\begin{equation}
    R^\star
    \sim C\frac{\delta}{|\varepsilon|}.
\end{equation}
At $\eta=0$ one obtains the cubic $\delta^{1/3}$ law. On the opposite side, $a\eta\to-\infty$, the largest positive root satisfies
\begin{equation}
    R_\eta
    \sim
    \sqrt{\frac{-a\eta}{\gamma_0}},
\end{equation}
which matches the fold-side law
\begin{equation}
\label{eq:fold-side-law}
    R^\star
    \sim
    \frac{\|v\|_\infty}{2}
    \sqrt{\frac{-a\varepsilon}{\gamma_0}}.
\end{equation}
These are outer matching limits of the profile $D(\eta)$; Theorem~\ref{thm:crossover} itself asserts uniformity only for $\eta$ in compact sets. Thus a pair of folds approaches the nominal state at distance $O(\sqrt{|\varepsilon|})$. For every fixed nonzero $\varepsilon$, the inverse function theorem still gives a sufficiently small regular neighborhood; the fold-side term appears because this neighborhood itself shrinks on the same scale as the reconstruction problem in the joint limit.

For measurement-time design, the result has two implications. First, a condition number based on $DF_\tau$ is not uniformly informative near rank-changing schedules. Second, the side from which a singular timing is approached matters: two schedules with the same $|\varepsilon|$ can have qualitatively different finite-noise geometry.

%% file: sections/05_benchmarks.tex
\section{Nonlinear Mechanical Benchmarks}
\label{sec:benchmarks}

\begin{figure*}[t]
    \centering
    \includegraphics[width=0.97\textwidth]{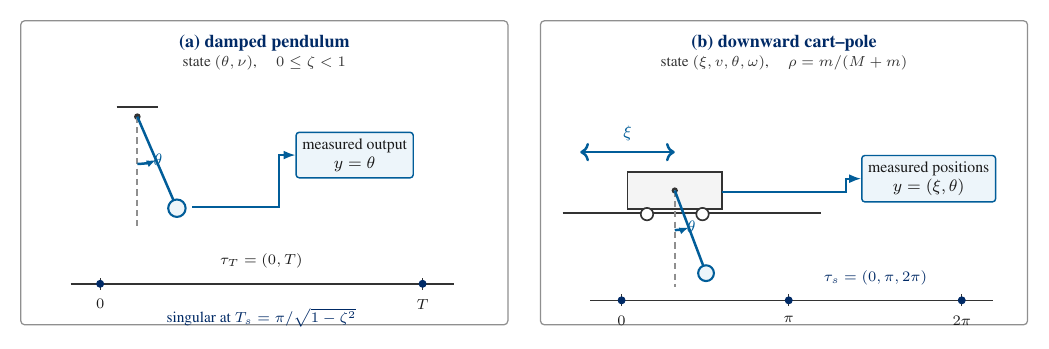}
    \caption{Nonlinear mechanical benchmarks and their singular sampling schedules. (a) The damped pendulum is measured through angle only at $\tau_T=(0,T)$, with the first singular schedule at $T_s=\pi/\sqrt{1-\zeta^2}$. (b) The downward cart--pole is measured through cart position and pendulum angle at $\tau_s=(0,\pi,2\pi)$, producing the rectangular sampled-output map used to test the corank-one cubic theory.}
    \label{fig:benchmark-models}
\end{figure*}

We use two benchmarks with different purposes. The damped pendulum provides a closed-form physical realization of the complete regular--singular--fold transition. The downward cart--pole is four-dimensional and produces an overdetermined sampled-output map, thereby testing the rectangular quotient formulation and the trajectory-sensitivity computation. Figure~\ref{fig:benchmark-models} summarizes the measured coordinates and singular sampling schedules used below.

\subsection{Damped pendulum: analytic singular timing and crossover}

Consider the dimensionless damped pendulum
\begin{equation}
\label{eq:damped-pendulum}
    \theta'=\nu,
    \qquad
    \nu'=-2\zeta\nu-\sin\theta,
    \qquad
    y=\theta,
\end{equation}
with $0\le\zeta<1$. Let
\begin{equation}
    \omega=\sqrt{1-\zeta^2}.
\end{equation}
For the two-sample schedule $\tau_T=(0,T)$, the linearized sampled Jacobian is
\begin{equation}
    J_T
    =
    \begin{bmatrix}
        1&0\\
        \phi_{11}(T)&\phi_{12}(T)
    \end{bmatrix},
\end{equation}
where
\begin{align}
    \phi_{11}(T)
    &=e^{-\zeta T}
    \left(
    \cos\omega T
    +\frac{\zeta}{\omega}\sin\omega T
    \right),\\
    \phi_{12}(T)
    &=e^{-\zeta T}\frac{\sin\omega T}{\omega}.
\end{align}
The first positive singular time is therefore
\begin{equation}
\label{eq:pendulum-Ts}
    \boxed{T_s=\frac{\pi}{\omega}.}
\end{equation}
Set
\begin{equation}
    \alpha=e^{-\zeta T_s}.
\end{equation}
At $T_s$,
\begin{equation}
    J_{T_s}
    =
    \begin{bmatrix}
        1&0\\
        -\alpha&0
    \end{bmatrix},
\end{equation}
so one may take
\begin{equation}
    v=(0,1)^T,
    \qquad
    \lambda=(\alpha,1)^T.
\end{equation}
The odd symmetry of \eqref{eq:damped-pendulum} implies $D^2F_{T_s}(0)=0$. For $0<\zeta<1$, direct third-order sensitivity calculation gives
\begin{equation}
\label{eq:pendulum-M}
    M_{T_s}
    =
    \frac{\alpha(1-\alpha^2)}
    {8\zeta(4-3\zeta^2)}
    \begin{bmatrix}
        2(1+6\zeta^2)&9\zeta\\
        9\zeta&6
    \end{bmatrix}.
\end{equation}
The inner matrix has determinant $3(4-3\zeta^2)>0$, hence
\begin{equation}
    M_{T_s}\succ0
    \qquad
    \forall\,0<\zeta<1.
\end{equation}
At $\zeta=0$, \eqref{eq:pendulum-M} is understood by continuous extension, yielding
\begin{equation}
    M_{T_s}
    \to
    \operatorname{diag}\left(\frac{\pi}{8},\frac{3\pi}{8}\right)
    \succ0.
\end{equation}
Equation~\eqref{eq:mu3-square} therefore yields, for $0<\zeta<1$,
\begin{equation}
\label{eq:pendulum-mu3}
    \boxed{
    \mu^{(3)}(T_s)
    =
    \frac{\alpha(1-\alpha)}
    {32\zeta(4-3\zeta^2)},}
\end{equation}
with the continuous undamped limit $\mu^{(3)}\to\pi/128$ as $\zeta\to0$.

Near the singular timing,
\begin{equation}
    \mu^{(1)}(T_s+\varepsilon)
    =
    \frac{\alpha}{1+\alpha}|\varepsilon|
    +O(\varepsilon^2).
\end{equation}
The signed transversal coefficient of Corollary~\ref{cor:scalar-crossover} is
\begin{equation}
    a=-\frac{\alpha}{1+\alpha}.
\end{equation}
Thus, with
\begin{equation}
    \eta=\frac{T-T_s}{\delta^{2/3}},
\end{equation}
the complete transition profile is the largest positive root of
\begin{equation}
\label{eq:pendulum-profile}
    \boxed{
    \mu^{(3)}(T_s)R^3
    -\frac{\alpha}{1+\alpha}\eta R
    =2.}
\end{equation}
For this orientation, $T<T_s$ is the regular side and $T>T_s$ is the fold side.

\begin{figure*}[t]
    \centering
    \includegraphics[width=0.97\textwidth]{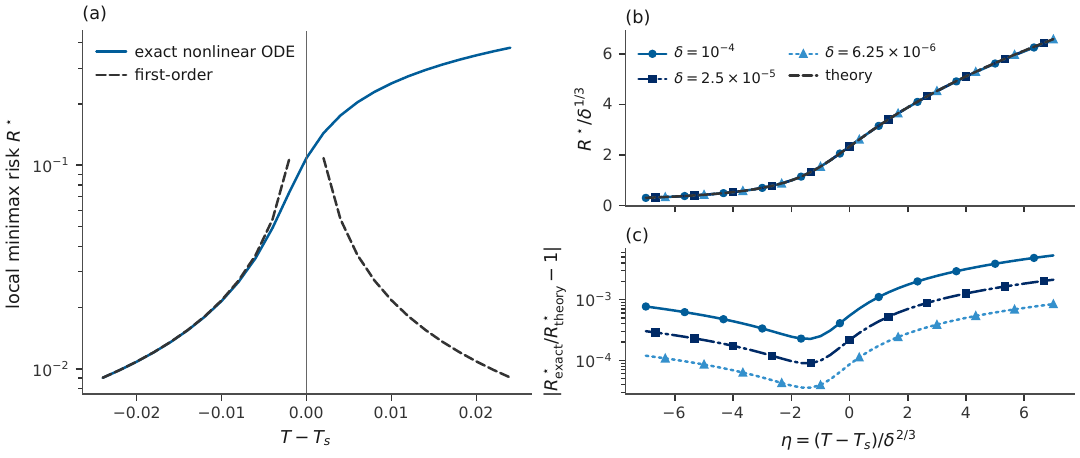}
    \caption{Damped-pendulum timing transition for $\zeta=0.05$. (a) Exact nonlinear local minimax risk versus timing offset for the representative noise level $\delta=10^{-4}$; the dashed curve is the first-order prediction. (b) Rescaled exact risks for $\delta=10^{-4}$, $2.5\times10^{-5}$ and $6.25\times10^{-6}$ collapse under $\eta=(T-T_s)/\delta^{2/3}$ onto the signed crossover profile. (c) Relative profile error $|R^\star_{\mathrm{exact}}/R^\star_{\mathrm{theory}}-1|$ on a logarithmic vertical axis, quantifying the deviations that are nearly indistinguishable in (b). Panels (b) and (c) share the noise-level colors, markers and line styles. All exact risks in this sweep lie below the theoretical prediction.}
    \label{fig:pendulum-transition}
\end{figure*}

\begin{figure}[t]
    \centering
    \includegraphics[width=\columnwidth]{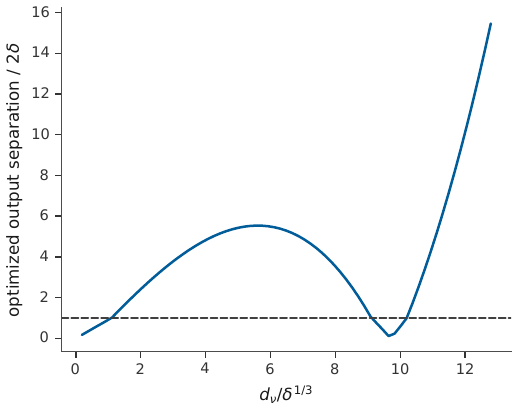}
    \caption{Fold-side pairwise geometry at $\delta=10^{-4}$ and $\eta=4$. The normalized optimized output separation is nonmonotone; the minimax diameter is determined by the largest feasible crossing of the threshold one.}
    \label{fig:pendulum-fold}
\end{figure}

The numerical study integrates the exact nonlinear ODE and solves the pairwise indistinguishability problem directly. For $\zeta=0.05$, the exact cubic-risk ratio converges to the theoretical coefficient as $\delta\downarrow0$, while plotting $R^\star/\delta^{1/3}$ against $\eta$ collapses different noise levels onto \eqref{eq:pendulum-profile}. Selected unrestricted pair optimizations agree with the symmetry-reduced solver on the regular, singular, and fold sides. Figure~\ref{fig:pendulum-transition} shows a representative timing transition at $\delta=10^{-4}$ in (a), the three-noise-level crossover collapse in (b), and the relative profile error in (c), which quantifies the deviations concealed by the nearly overlapping curves in (b), while Fig.~\ref{fig:pendulum-fold} exposes the nonmonotone fold-side separation geometry that makes the largest feasible crossing relevant.

\subsection{Four-state downward cart--pole: rectangular cubic geometry}

We consider the stable downward operating regime of the standard cart--pole mechanism. The same mechanical apparatus is commonly operated about the upright equilibrium as an inverted pendulum on a cart; here the downward equilibrium is used because it provides an oscillatory nonlinear benchmark with an analytically accessible singular sampling schedule.

Let
\begin{equation}
    \rho=\frac{m}{M+m}\in(0,1)
\end{equation}
be the pendulum-to-total mass ratio. Normalize cart displacement by the pendulum length and time by the downward small-signal frequency. With state
\begin{equation}
    X=(\xi,v,\theta,\omega),
\end{equation}
the exact unforced dynamics are
\begin{align}
    \xi'&=v,\\
    v'&=
    \frac{\rho\sin\theta}
    {1-\rho\cos^2\theta}
    \left[
    \omega^2+(1-\rho)\cos\theta
    \right],\\
    \theta'&=\omega,\\
    \omega'&=
    -\frac{\sin\theta}
    {1-\rho\cos^2\theta}
    \left[
    (1-\rho)+\rho\omega^2\cos\theta
    \right].
\end{align}
We measure both positions,
\begin{equation}
    y=(\xi,\theta),
\end{equation}
at
\begin{equation}
\label{eq:cartpole-sing-schedule}
    \boxed{\tau_s=(0,\pi,2\pi).}
\end{equation}
Thus $F_{\tau_s}:\R^4\to\R^6$.

The linearization is
\begin{equation}
    \xi''=\rho\theta,
    \qquad
    \theta''=-\theta,
\end{equation}
with
\begin{equation}
    \xi(s)
    =
    \xi_0+v_0s
    +\rho\theta_0(1-\cos s)
    +\rho\omega_0(s-\sin s).
\end{equation}
At \eqref{eq:cartpole-sing-schedule}, the two initial velocities appear only through $v_0+\rho\omega_0$. Hence
\begin{equation}
    \rank J_s=3,
\end{equation}
with null direction
\begin{equation}
    \boxed{
    v_s=(0,\rho,0,-1)^T.}
\end{equation}

The downward cart--pole has an exact common-shift symmetry in cart position and velocity. Adding the same $(a,b,0,0)$ to both candidate initial states shifts every cart-position sample by $a+bt$ and leaves the pendulum motion unchanged. Pairwise output differences are therefore invariant under
\begin{equation}
    \mathcal S=\Span\{e_\xi,e_v\}.
\end{equation}
Following the symmetry reduction underlying Assumption~\ref{ass:cubic-coercivity}, the effective midpoint variables are the pendulum components $(\theta,\omega)$.

For $\rho=1/6$, direct integration of the third trajectory sensitivities simplifies the quotient cubic jet to
\begin{equation}
\label{eq:cartpole-gamma3}
    \boxed{
    \|\Gamma_3(\theta,\omega)\|_{J_s,\infty}
    =
    \frac{9\pi}{80}
    \left(
        \theta^2+3\omega^2+\frac14
    \right).}
\end{equation}
Thus the effective midpoint dependence is coercive and is uniquely minimized at the centered pair. Since $\|v_s\|_\infty=1$, Theorem~\ref{thm:rectangular-cubic} gives the exact modulus
\begin{equation}
\label{eq:cartpole-mu3}
    \boxed{
    \mu^{(3)}(\tau_s)
    =\frac{9\pi}{320}
    \approx0.08835729338.}
\end{equation}

As an independent check, construct the asymptotically minimizing pair using the null direction and the $O(d^3)$ regular correction returned by the quotient LP. Direct integration of the exact nonlinear downward cart--pole gives the ratios
\begin{equation}
\begin{split}
    &0.0885498,\ 0.0883955,\ 0.0883662,\\
    &0.0883595,\ 0.0883578
\end{split}
\end{equation}
for $d=0.4,0.2,0.1,0.05,0.025$, respectively, converging to \eqref{eq:cartpole-mu3}.

\begin{figure*}[t]
    \centering
    \includegraphics[width=0.97\textwidth]{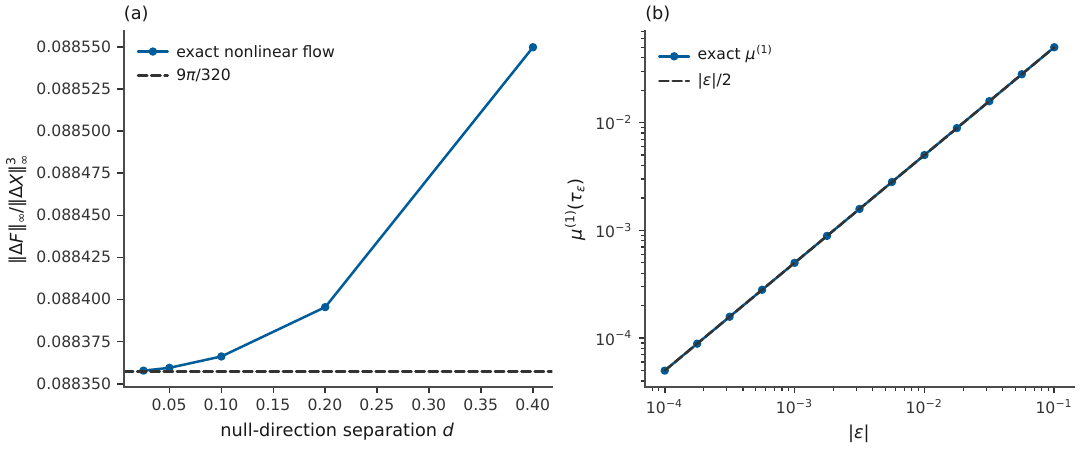}
    \caption{Four-state downward cart--pole validation for $\rho=1/6$. (a) Exact nonlinear cubic ratio converging to the analytic modulus $9\pi/320$. (b) Restoration of the physical $\ell_\infty$ first-order margin as the last sampling time moves away from $2\pi$; the dashed line is $|\varepsilon|/2$.}
    \label{fig:cartpole-validation}
\end{figure*}

Finally, perturb only the last measurement time,
\begin{equation}
    \tau_\varepsilon=(0,\pi,2\pi+\varepsilon).
\end{equation}
For $\varepsilon\ne0$ the sampled Jacobian regains full column rank, and exact $\ell_\infty$ LP evaluation gives
\begin{equation}
    \boxed{
    \mu^{(1)}(\tau_\varepsilon)
    \sim\frac{|\varepsilon|}{2}.}
\end{equation}
Figure~\ref{fig:cartpole-validation} shows both the exact nonlinear convergence to the analytic cubic coefficient and the linear restoration of the physical first-order margin. The downward cart--pole therefore verifies the complete computational workflow on a four-state, overdetermined sampled-output map.

%% file: sections/06_conclusion.tex
\section{Conclusion}
\label{sec:conclusion}

A rank-deficient sampled Jacobian does not by itself determine the bounded-noise reconstruction behavior of a nonlinear system. This paper develops a finite-type minimax description in which a measurement schedule is characterized locally by a robust reconstruction order $k(\tau)$ and a leading physical-norm modulus $\mu^{(k)}(\tau)$. Regular schedules recover the first-order minimum gain, whereas cubic corank-one schedules lead to a $\delta^{1/3}$ law that can be computed from second and third trajectory sensitivities, including for overdetermined sampled-output maps.

The timing analysis shows that the transition through a cubic singular schedule is inherently joint in timing and noise. The scale $|\varepsilon|=O(\delta^{2/3})$ produces a uniform signed crossover: one side approaches the ordinary regular law, while the other develops nearby folds whose separation is already visible at the reconstruction-error scale. The damped-pendulum and cart--pole examples confirm these mechanisms on exact nonlinear dynamics.

The results suggest that measurement-time design near rank-changing schedules should not be based solely on first-order conditioning. A natural next direction is to extend the finite-type geometry beyond a one-dimensional Jacobian kernel, where different invisible directions may become observable at different nonlinear orders.

%% file: sections/appendix_proofs.tex
\section{Proof of the Rectangular Cubic Modulus}
\label{app:rectangular-proof}

We record the main steps required for Theorem~\ref{thm:rectangular-cubic}; the argument is organized around the exact reduced coordinates and the physical quotient norm used in the theorem statement.

Let $\xi=(w,z)$ denote the exact reduced coordinates and write
\begin{equation}
    Q_z(\xi)=R_2(\xi)+E_3(\xi),
    \qquad
    \|E_3(\xi)\|_{J,\infty}\le C\|\xi\|^3.
\end{equation}
The mixed condition \eqref{eq:mixed-obstruction} is exactly the statement that the linear term of $Q_z$ vanishes in the quotient.

For two reduced states whose singular-coordinate difference is $d$, write their common midpoint as $\bar\xi$. If an exact common-shift symmetry is present, shift both physical states so that $\bar\xi$ lies in a fixed transversal complement. This leaves both state and output pairwise differences unchanged.

First consider a fixed scaled midpoint $c$ and the pair
\begin{equation}
    \xi_\pm
    =d\left(c\pm\frac12e_n\right).
\end{equation}
The quotient output increment is
\begin{equation}
    \pi_J(F_+-F_-)
    =d^3\Gamma_3(c)+O(d^4).
\end{equation}
An $O(d^3)$ correction in the regular physical state directions changes the leading output by an arbitrary element of $\im J$ without changing the leading state separation. Hence one can realize the quotient distance and obtain
\begin{equation}
    \limsup_{d\to0}
    \frac{\|F_+-F_-\|_\infty}
    {\|x_+-x_-\|_\infty^3}
    \le
    \frac{\|\Gamma_3(c)\|_{J,\infty}}
    {\|v\|_\infty^3}.
\end{equation}
Taking the infimum over $c$ gives the upper bound in \eqref{eq:mu3-rectangular}.

For the lower bound, consider any sequence of pairs with bounded cubic output ratio. Exact regular coordinates imply
\begin{equation}
    \Delta w=O(r^3),
\end{equation}
where $r$ is the physical state separation. Local bi-Lipschitz equivalence then gives
\begin{equation}
    |d|\asymp r,
    \qquad
    \Delta x=dv+o(d).
\end{equation}
It remains to control the midpoint. Assumption~\ref{ass:cubic-coercivity} implies quadratic coercivity of the homogeneous part on the effective midpoint space:
\begin{equation}
    \|R_2(c)\|_{J,\infty}
    \ge m\|c\|^2.
\end{equation}
Using the fundamental theorem of calculus across the singular coordinate,
\begin{equation}
    \Delta Q
    =d\int_{-1/2}^{1/2}
    Q_z(\bar\xi+ud e_n)\,du.
\end{equation}
If $\|\bar\xi\|/|d|\to\infty$, the leading term is bounded below by a constant multiple of
\begin{equation}
    |d|\|\bar\xi\|^2,
\end{equation}
whereas the Taylor remainder is
\begin{equation}
    O\left(
    |d|\|\bar\xi\|^3+|d|^4
    \right).
\end{equation}
Since $\bar\xi\to0$, this contradicts bounded cubic output ratio. Thus $c=\bar\xi/d$ remains bounded. The uniform expansion
\begin{equation}
    \pi_J(\Delta F)
    =d^3\Gamma_3(c)+o(d^3)
\end{equation}
now applies, yielding
\begin{equation}
    \|\Delta F\|_\infty
    \ge
    |d|^3\chi_3^{\rm rect}+o(|d|^3).
\end{equation}
Together with $r=|d|\|v\|_\infty+o(|d|)$, this gives the matching lower bound and proves Theorem~\ref{thm:rectangular-cubic}.

\section{Proof of the Uniform Crossover Theorem}
\label{app:crossover-proof}

Set
\begin{equation}
    \varepsilon=\eta\delta^{2/3},
\end{equation}
with $\eta$ restricted to a compact set $K$. All candidate pairs lie in $X_\delta=B_\infty(0,\rho(\delta))$, so their endpoints converge to the nominal state. Conversely, because $\delta^{1/3}=o(\rho(\delta))$, every bounded scaled transition-layer construction remains inside $X_\delta$ for sufficiently small $\delta$. The expansion \eqref{eq:crossover-local-expansion} is therefore uniform on the relevant pairs.

For the lower bound, choose $R<R_\star(\eta)$ with $G_\eta(R)<2$ and choose $c\in\mathcal N_\xi$ such that
\begin{equation}
    \|\eta R A+R^3\Gamma_3(c)\|_{0,\infty}<2.
\end{equation}
The boundary-approach condition \eqref{eq:boundary-approach} allows such $R$ to approach $R_\star(\eta)$ arbitrarily closely. Let
\begin{equation}
    d=R\delta^{1/3},
    \qquad
    \bar\xi=dc.
\end{equation}
Then
\begin{equation}
    \pi_0(\Delta F)
    =
    \delta\left[
    \eta RA+R^3\Gamma_3(c)
    \right]
    +o_K(\delta).
\end{equation}
As in Appendix~\ref{app:rectangular-proof}, an $O(\delta)$ regular state correction realizes the quotient distance in the physical output norm. Hence such pairs are indistinguishable for sufficiently small $\delta$, with state separation
\begin{equation}
    \|\Delta x\|_\infty
    =
    R\|v\|_\infty\delta^{1/3}
    +o_K(\delta^{1/3}).
\end{equation}
Letting $R\uparrow R_\star(\eta)$ yields the lower diameter bound $D(\eta)\delta^{1/3}$.

For the upper bound, take any sequence $\delta_k\downarrow0$, $\eta_k\in K$, and indistinguishable pairs in $X_{\delta_k}$. If a family-wide exact common-shift symmetry is present, first shift both endpoints to the fixed transversal midpoint slice; pairwise state and output differences are unchanged. We show that their state separation satisfies
\begin{equation}
    r_k=O(\delta_k^{1/3})
\end{equation}
uniformly for $\eta_k\in K$. Otherwise $|d_k|/\delta_k^{1/3}\to\infty$. After replacing the two regular endpoint coordinates by their midpoint, the quotient output has leading form
\begin{equation}
    \varepsilon_k d_k A+d_k^3\Gamma_3(c_k).
\end{equation}
Because
\begin{equation}
    \frac{\varepsilon_k}{d_k^2}
    =
    \eta_k\frac{\delta_k^{2/3}}{d_k^2}
    \to0,
\end{equation}
cubic coercivity forces $\|\Delta F_k\|_\infty\gg\delta_k$, contradicting indistinguishability.

Thus $R_k=|d_k|/\delta_k^{1/3}$ is bounded. If $R_k\to0$, the corresponding state separation is $o(\delta_k^{1/3})$ and does not affect the leading diameter coefficient. Otherwise, after passing to a subsequence, $R_k\to R>0$, the scaled midpoint $c_k$ is bounded by the same coercivity argument, and $\eta_k\to\eta\in K$. The normalized quotient increment converges to
\begin{equation}
    \eta RA+R^3\Gamma_3(c).
\end{equation}
Indistinguishability implies
\begin{equation}
    G_\eta(R)\le2,
\end{equation}
so $R\le R_\star(\eta)$ and therefore
\begin{equation}
    \limsup_{k\to\infty}
    \delta_k^{-1/3}r_k
    \le D(\eta).
\end{equation}
Continuity of $R_\star$ on $K$ upgrades this subsequence argument to uniform convergence. Proposition~\ref{prop:pairwise} then supplies the factor $1/2$ and proves Theorem~\ref{thm:crossover}.